\documentstyle[12pt,aaspp4]{article}

\def \ref {\setlength{\parskip}{0pt}\noindent\hangindent=0.5in\hangafter=1}

\begin{document}

\vspace*{-1.0in}
{\center{Accepted for publication in the ApJ Letters HST Second Servicing Mission special issue}}

\title {The STIS Parallel Survey: Introduction and First
Results\footnote[1]{ Based on observations made with the NASA/ESA
{\em Hubble Space Telescope}, obtained from the data archive at the Space
Telescope Science Institute, which is operated by the Association
of Universities for Research in Astronomy, Inc., under NASA contract
NAS 5-26555.}.}

\author {Jonathan P. Gardner\altaffilmark{2,8}, 
Robert S. Hill\altaffilmark{2,3},
Stefi A. Baum\altaffilmark{4},
Nicholas R. Collins\altaffilmark{2,3},
Henry C. Ferguson\altaffilmark{4},
Robert A. E. Fosbury\altaffilmark{5},
Ronald L. Gilliland\altaffilmark{4},
Richard F. Green\altaffilmark{6},
Theodore R. Gull\altaffilmark{2},
Sara R. Heap\altaffilmark{2},
Don J. Lindler\altaffilmark{2,7},
Eliot M. Malumuth\altaffilmark{2,3},
Alberto Micol\altaffilmark{5},
Norbert Pirzkal\altaffilmark{5},
Jennifer L. Sandoval\altaffilmark{2,7},
Eline Tolstoy\altaffilmark{5},
Jeremy R. Walsh\altaffilmark{5},
Bruce E. Woodgate\altaffilmark{2}
}

\altaffiltext{2}{Laboratory for Astronomy and Solar Physics,
Code 681, Goddard Space Flight Center, Greenbelt MD 20771 }
\altaffiltext{3}{ Hughes STX Corp., Lanham MD 20706 }
\altaffiltext{4}{Space Telescope Science Institute,
3700 San Martin Drive, Baltimore MD 21218 }
\altaffiltext{5}{ Space Telescope--European Coordinating Facility, Karl
Schwarzschild Str. 2, D-85748, Garching bei M\"{u}nchen, Germany }
\altaffiltext{6}{ National Optical Astronomy Observatories, P.O. Box 26732, 
Tucson AZ 85726 }
\altaffiltext{7}{ Advanced Computer Concepts, Inc., 11518 Gainsborough Road,
Potomac MD 20854 }
\altaffiltext{8}{NOAO Research Associate \\
\\
{E-mail addresses: gardner@harmony.gsfc.nasa.gov,
bhill@virgil.gsfc.nasa.gov, sbaum@stsci.edu, \\
collins@zolo.gsfc.nasa.gov,
ferguson@stsci.edu, rfosbury@eso.org, gillil@stsci.edu, 
green@noao.edu, \\
gull@sea.gsfc.nasa.gov, hrsheap@stars.gsfc.nasa.gov,
lindler@rockit.gsfc.nasa.gov, \\
eliot@barada.gsfc.nasa.gov, 
amicol@eso.org,
npirzkal@eso.org, sandoval@hires.gsfc.nasa.gov, \\
etolstoy@eso.org,
jwalsh@eso.org, woodgate@stars.gsfc.nasa.gov}}

\begin{abstract}

The installation of the Space Telescope Imaging Spectrograph (STIS)
on the Hubble Space Telescope (HST) allows for the first time
two-dimensional optical and ultraviolet slitless spectroscopy of
faint objects from space. The STIS Parallel Survey (SPS) routinely
obtains broad band images and slitless spectra of random fields in
parallel with HST observations using other instruments. The SPS is
designed to study a wide variety of astrophysical phenomena,
including the rate of star formation in galaxies at intermediate
to high redshift through the detection of emission-line galaxies.
We present the first results of the SPS, which demonstrate the
capability of STIS slitless spectroscopy to detect and identify
high-redshift galaxies.

\end{abstract}

%\keywords{
%cosmology: observations ---
%galaxies: evolution ---
%galaxies: luminosity function, mass function ---
%galaxies: statistics ---
%surveys
%}

\section{Introduction}

The low sky background seen by the Hubble Space Telescope (HST),
and the two-dimensional capability of the Space Telescope Imaging
Spectrograph (STIS; Kimble et al.\ 1997) enables a survey of faint
spectroscopically selected objects for the first time. The STIS
Parallel Survey (SPS) uses slitless spectroscopy of random fields
to identify and study objects selected by their spectroscopic
properties. STIS slitless spectra provide unprecedented sensitivity
in the range $7000 < \lambda < 10000${\AA}, where ground based
spectroscopy is difficult due to the variability of the night-sky
lines. SPS observations include, as a fiducial for the spectra,
unfiltered images of each field, utilising between 1/6 and 1/4 of
the exposure time. The goals of the SPS include studying the
evolution of the star formation rate with redshift as measured by
emission-line strengths, the evolution of the luminosity function
of galaxies, the size and morphological evolution of galaxies with
redshift, the study of active galactic nuclei (AGN), weak gravitational
lensing, stellar luminosity functions in nearby galaxies, low-mass
stars and Kuiper Belt objects. The images will be used to construct
magnitude-limited samples and to identify the morphological types
of the galaxies in the spectroscopically selected samples. SPS data
are also a valuable source of calibration information for the STIS
CCD detector.

The SPS is being conducted as a service to the astronomical community
by the Space Telescope Science Institute (STScI), and the data are
made available immediately through the archive. Further information
is available on the world wide web at the STIS Investigation
Definition Team (IDT) parallel page, {\em
http://hires.gsfc.nasa.gov/stis/parallels/parallels.html}, the
STScI parallel page, {\em
http://www.stsci.edu/ftp/instrument\_news/STIS/parallel/parallelstis.html},
and the Space Telescope -- European Coordinating Facility parallel
page, {\em http://ecf.hq.eso.org/parallel}. In this {\em Letter}
we describe the survey data, demonstrate the capabilities of
random-field slitless spectroscopy with STIS, and show results from
some of the first fields to be observed, with an emphasis on star
formation at high redshift.

The observational study of galaxy evolution is the study of star
formation as a function of lookback time. Determining the star-formation
history of the universe, its variation with galaxy morphological
type, and its relationship to spectral type and color, places strong
constraints on theoretical models. Ground-based imaging and redshift
surveys, in combination with the HST Medium Deep Survey key project
(MDS; Griffiths et al.\ 1994), and the Hubble Deep Field (HDF;
Williams et al.\ 1996), have built up a consistent picture of evolution
in which the population of early-type galaxies have undergone little
star-formation at redshift $z<1$ (Lilly et al.\ 1995), while a population
of morphologically complex and active star-forming galaxies were
numerous at $z \sim 0.4$ (Cowie, Songaila \& Hu 1991; Glazebrook et al.\ 1995; Driver, Windhorst \& Griffiths 1995),
but have disappeared by the present epoch. Star-forming galaxies
has recently been discovered at $z>3$ (Steidel et al.\ 1996), but the
relatively small number of ``UV-dropout'' galaxies in the HDF
indicates that the bulk of star formation takes place at $z<3$
(Madau et al.\ 1996). Although there is still considerable debate about
this picture, it is clear that the poorly studied redshift range
$1<z<3$ is of considerable importance.

Galaxy formation may not be a dramatic process, but might be a slow
and steady accumulation of hierarchical merging and constant star
formation (Cole et al.\ 1994). Lanzetta, Yahil \& Fernandez-Soto (1996) argue that many of
the galaxies in the HDF are at $z<3$, and Cowie et al.\ (1988) and
Gardner (1995) argue on the basis of surface brightness considerations
that the star-formation which produced the locally-observed metals
occurred in galaxies already detected in the deep photometric
surveys. While the small scatter in the color of cluster elliptical
galaxies has normally been taken to mean a common epoch of formation,
and thus an old age, recent theoretical work has managed to reproduce
this observation with a hierarchical merging model (Baugh et al.\ 1996).
In such a picture, the redshift range of greatest interest might
be $1<z<3$ rather than higher redshifts. SPS observations are
sensitive to H$\alpha$ emission at $z<0.5$, [O~{\sc iii}] emission
at $z<1.0$, [O~{\sc ii}] emission at $0.5<z<1.5$ and Ly$\alpha$
emission at $3.5<z<7.0$.

While searches for Ly$\alpha$ emission from forming galaxies have
for the most part been unsuccessful (for a review see Pritchet 1994),
there are several reasons the SPS might be at an advantage. Primeval
galaxies may not emit much Ly$\alpha$, due to partial obscuration
by dust, although Pritchet (1994) calculates that $\ge 10\%$ of
primeval galaxies should be visible in Ly$\alpha$, a percentage
which is roughly consistent with the observations of Steidel et al.\ (1996).
Primeval galaxies may be large, low surface-brightness objects,
but the high resolution of the HST may be able to detect individual
regions of star-formation (i.e. H~{\sc ii} regions) within a larger
low-surface brightness galaxy, Dalcanton \& Schectman (1996). It is also
possible that biasing, or strong clustering of galaxies makes
individual pencil-beam surveys of random fields a hit-or-miss
proposition.

\section{The SPS Data}

The STIS Parallel Survey began on 1997 June 2. The HST instruments
are distributed around the focal plane of the Optical Telescope
Assembly (OTA) and a primary instrument aperture is selected by
offsetting the pointing of the OTA from the optical axis. The
remaining instruments then view a random field in the sky between
5 and 8 arcminutes away from the primary object. Prior to the
installation of STIS and the Near Infrared Camera and Multi-Object
Spectrometer (NICMOS) during the second servicing mission in 1997
February, parallel observations were only scientifically useful
when taken with the Wide Field and Planetary Camera 2 (WFPC2). The
MDS, along with the Guaranteed Time Observer's parallel survey,
has used the WFPC2 in parallel to study many aspects of cosmology
and stellar populations, obtaining data in about 600 orbits per
year. Now, with three cameras capable of making parallel observations,
and the greater flexibility provided by the installation of a higher
storage capacity Solid State Recorder, the opportunity for random
field surveys has increased by a factor of 6 or more.

``Scripted'' parallel observations, in which the choice of exposure
times, filters and spectroscopic modes depends on the Galactic
latitude or the available exposure time, will not be implemented
for scheduling the STIS until late in 1997. The observations made
to date have been in a ``non-scripted'' mode with two $150 s$
images, and two $600 s$ to $900 s$ spectroscopic exposures per
orbit, repeated for multiple orbit pointings. The images are taken
in the 50CCD clear camera mode, and are sensitive to $2200 < \lambda
< 11000${\AA} wavelengths (see Baum et al.\ 1996). The spectral images
use the G750L low resolution grating, with central wavelength
8975{\AA}, and have spectra covering about 4000{\AA} between 5500
and 11490{\AA}, depending on the position of the objects on the
field. The images are read out in a $2 \times 2$ binned pixel mode
while the spectra are binned by two pixels in the spatial direction
only, resulting in $0.1\arcsec$ pixels. SPS images taken after 1997
August, will be unbinned to make the data more suitable for weak
lensing measurements, and other scientific programs. The central
wavelength of the spectra will be changed to 7751{\AA} to take
greater advantage of the sensitivity of the CCD, which drops rapidly
at wavelengths longer than 10000{\AA}. At the time of writing, 7
weeks after the start of observations, 125 fields have been observed,
with spectroscopic exposure times up to $16800 s$. Table~1 lists
the longest exposures made to date, and the full list is available
on the STIS IDT web page.

\subsection{Data Reduction}

The data reduction of SPS data has three main stages: ({\em i})
image reduction, including bias and dark subtraction, hot pixel
correction and flat fielding; ({\em ii}) co-addition of images from
multiple and dithered pointings; ({\em iii}) object detection and
slitless spectrum extraction. We implement these procedures in the
Interactive Data Language (IDL), except for the object detection,
which is done using the SExtractor package (Bertin \& Arnouts 1996).

If the prime observations are dithered, the parallel observations
will also be dithered. Visits to the same prime target done at
different times can result in offsets between parallel observations
of the same field due to change in the roll angle of the spacecraft.
The SPS data are put into groups within $5\arcsec$ of each other
($\sim 10\%$ of the size of the STIS CCD), and subgroups within
$0.025\arcsec$ of each other. The images in each subgroup are
treated as co-pointed for the purpose of cosmic ray (CR) removal.

Cosmic rays (CRs) are removed using multiple exposures of the same
fields. A one-dimensional bias is computed from the overscan portion
of the detector and subtracted from the raw image. Two-dimensional
bias and dark images are subtracted. Some pixels with high dark
rates are not accounted for by the dark frame, since new ``hot
pixels'' are continually generated by the CR flux on orbit, and
periodically reduced by annealing. These pixels are tabulated
weekly. The hot pixels corrected in the data reduction are the
union of the latest list made before the given image and the earliest
list made afterward. The correction is done by linear interpolation
between neighboring pixels. The images are flat fielded using a
median sky flat of the SPS images. Dithered data are co-registered,
and the resulting images are co-added by a final step which rejects
CRs and hot pixels using all the available data. The processing
for slitless spectra is similar to what is described above, except
that the dither offsets are computed from those found for the direct
images, the spectra are not flat fielded, and a background determined
by the median of each column is subtracted. No attempt is made to
remove the fringing.

The program SExtractor (Bertin \& Arnouts 1996) detects and separates sources
on the processed images using a multi-thresholding algorithm.
SExtractor subtracts a smoothed 2-dimensional background and
convolves the image with a $0.5\arcsec \times 0.5\arcsec$ Gaussian
filter before applying the initial detection threshold, which is
$0.7\sigma$ of the empirical sky noise. The program computes
location, photometry, and shape parameters for each detected source.

Slitless spectra are extracted as rectangular sub-images, which are
resampled using bilinear interpolation to remove geometric distortion.
An array of pixel offsets characterizing the distortion is maintained
as part of the local calibration database. The first axis of the
spectrum image is wavelength, with the zero point and dispersion
determined from the object's location in the camera mode image.
The second axis is spatial position in the cross-dispersion direction,
corrected for the orientation angle of the object. The rectified
image is then summed in the cross-dispersion direction. In this
{\em Letter} we used unweighted summation over a simple rectangular
extraction slit. However, a scheme for weighted, or ``optimal,''
extraction has recently been implemented (Horne 1986). Optimal
extraction increases the signal-to-noise ratio by weighting pixels
by their contribution to the total signal, and helps in rejection
of bad pixels.

Photometric calibration of the direct images was done independently
by three of the authors (EMM, ET and JW) using STIS 50CCD observations
of $\omega$ Centauri and the white dwarf, GRW+70D5824. The field
in $\omega$ Centauri has been observed with WFPC2 by Holtzman et al.\ (1995)
and from the ground by Harris et al.\ (1993), and the photometry was
determined relative to those observations.

\section{Discussion}

A preliminary analysis of the SPS data has discovered 9
emission-line sources, ranging in redshift $0.12<z<0.81$.
Figure~1 shows the spectra, with the strongest
lines marked. For each object, the extracted spectrum is shown,
and the spectrum smoothed with a Gaussian filter three pixels wide is
shown offset upwards for clarity. Table~2 contains additional
information about the detected emission line objects. These objects
all show two or more emission lines; in addition the data contain
several single emission-line objects which will require additional
observations to confirm their redshifts. These objects could be at
$1<z<1.5$ if the single emission line is [O~{\sc ii}] at rest
wavelength 3727{\AA}.

SPS1514+3632\#008 is our strongest detection, and the spectrum is
plotted at the top of Figure~1. The [O~{\sc iii}] doublet is
clearly seen at 9018{\AA} and 8935{\AA}, indicating a redshift of
0.801. The H$\beta$ line is also seen at the same redshift. As a
demonstration of slitless spectroscopy with STIS, we show this
field in Figure~2, Plate~1. At the top right is the image, with a
total exposure of 2100 seconds, and a limiting detection magnitude
($5\sigma$ in a 0.5 arcsecond aperture) of AB = 28.0 mag. The bottom
of the figure is the slitless spectral image, with a total exposure
time of $8400 s$. At the top left of the figure we show an expanded
view of the spectral image in the region around the detected emission
lines. The object could be part of an interacting pair, and both
members of the pair are visible in the stronger [O~{\sc iii}] line
at 5007{\AA}.

\section{Summary}

The STIS Parallel Survey is designed to address a wide variety of
topics in astronomy, including the discovery and study of emission
line objects at redshift $z \approx 1$. The STIS provides unprecedented
sensitivity in the wavelength range $7000 < \lambda < 10000${\AA}
due to the low and constant sky background relative to observations
made from the ground. In the first seven weeks of operation, the
SPS has obtained images and slitless spectra of 125 fields.  We
have analyzed nine fields at high Galactic latitude with spectroscopic
exposure times of $5400 s$ or more. We find nine emission-line
objects with $0.12<z<0.81$. While it is still too early for a
statistical analysis of the data, it is clear that this survey will
contribute to our understanding of star formation at one half the
Hubble time.

We wish to acknowledge useful discussions with Ray Weymann, Thomas
Erben, Mario Nonino, Michael Rosa and Benoit Pirenne, and support and
help from the STIS Investigation Definition Team. We thank Karl
Glazebrook for providing redshift identification software.

\clearpage

{\scriptsize
\begin{deluxetable}{lrrrrrrrc}

\tablecaption{The longest SPS observations in the first 7 weeks of
operations}

\tablehead{
\colhead{Field} &
\colhead{Rootname} &
\colhead{Date} &
\colhead{RA (2000)} &
\colhead{Dec} &
\colhead{Galactic b} &
\colhead{Exp(image)} &
\colhead{Exp(spect)} &
}

\startdata
SPS1312$-$0124&{\tt O4141R020}&22/06/97&13:11:45.06&$-$01:24:05.1&61.05&2850&16800&*\nl
SPS1229+1245&{\tt O41Q70HBM}&15/06/97&12:28:40.42&+12:45:18.8&74.67&3300&13800&*\nl
SPS1550+2122b&{\tt O41Q19WSM}&10/06/97&15:49:35.19&+21:21:43.7&49.19&2100&11400&*\nl
SPS1550+2122a&{\tt O41467020}&05/06/97&15:49:37.17&+21:21:46.5&49.18&1800&10800&*\nl
SPS0020+5922&{\tt O41499020}&08/06/97&00:19:58.94&+59:21:34.0&$-$3.27&1800&10800&\nl
SPS1150+1241&{\tt O41406020}&09/06/97&11:49:56.56&+12:41:32.0&69.61&1500&9000&\nl
SPS0535$-$6918a&{\tt O41420020}&10/07/97&05:34:35.97&$-$69:17:43.2&$-$32.01&1500&9000&\nl
SPS1514+3632&{\tt O41Q24J2M}&16/06/97&15:14:21.02&+36:31:34.5&58.43&2100&8400&*\nl
SPS0039+4924&{\tt O41410020}&09/06/97&00:38:44.24&+48:23:33.1&$-$14.43&1350&7800&\nl
SPS1025+4703&{\tt O4141C020}&29/06/97&10:25:17.00&+47:02:40.9&55.18&1350&7800&\nl
SPS0039+4827&{\tt O41492020}&03/06/97&00:38:57.51&+48:27:04.1&$-$14.37&1200&7200&\nl
SPS1327$-$4737&{\tt O41485020}&04/06/97&13:26:35.85&$-$47:36:42.2&14.84&1200&7200&\nl
SPS1150+1242&{\tt O41474020}&05/06/97&11:49:58.49&+12:42:19.4&69.63&1200&7200&*\nl
SPS1327$-$4732&{\tt O41428020}&11/06/97&13:27:15.37&$-$47:31:58.2&14.90&1200&7200&\nl
SPS1231+1214&{\tt O42R05E3M}&14/07/97&12:31:03.69&+12:14:26.2&74.36&1800&7200&\nl
SPS1418+5217&{\tt O42R11H2M}&27/06/97&14:18:09.54&+52:17:07.3&60.07&1500&6000&*\nl
SPS0134+3041&{\tt O42R1YBPM}&11/07/97&01:34:10.67&+30:40:38.2&$-$31.30&1050&6000&\nl
SPS1741$-$5340&{\tt O41482020}&04/06/97&17:41:18.99&$-$53:39:52.3&$-$12.03&900&5400&\nl
SPS1741$-$5338&{\tt O41479020}&05/06/97&17:41:18.55&$-$53:38:26.3&$-$12.02&900&5400&\nl
SPS0040+4144&{\tt O41460020}&14/06/97&00:40:00.68&+41:43:57.7&$-$21.09&900&5400&\nl
SPS1418+5223&{\tt O42R06EFM}&23/06/97&14:17:59.52&+52:22:56.7&60.01&1350&5400&*\nl
SPS1257+0423&{\tt O4140C020}&30/06/97&12:57:14.98&+04:23:11.2&67.22&900&5400&*\nl
SPS0047$-$2514&{\tt O4141X020}&09/07/97&00:47:19.69&$-$25:13:30.8&$-$87.89&900&5400&\nl
SPS0535$-$6918b&{\tt O4142H020}&12/07/97&05:34:35.55&$-$69:17:36.2&$-$32.01&900&5400&\nl

\tablecomments{Fields marked with a * are discussed in this paper. The
rootname is the designation of the first exposure of the field in the
STScI archive. Exposure times for the images and spectroscopy are in
seconds.}

\enddata
\end{deluxetable}
}

{\scriptsize
\begin{deluxetable}{lrrrrl}

\tablenum{2}

\tablecaption{Emission line galaxies detected by the SPS}

\tablehead{
\colhead{Object} &
\colhead{RA (2000)} &
\colhead{Dec} &
\colhead{ABmag} &
\colhead{z} &
\colhead{lines}
}

\startdata
SPS1418+5223\#002 & 14:18:00.69 & 52:22:39.3 & 20.45 & 0.278 & H$\alpha$, [S~{\sc ii}], [O~{\sc i}] \nl
SPS1312$-$0124\#004 & 13:11:45.60 & $-$02:35:35.1 & 20.44 & 0.383 & H$\alpha$, [O~{\sc i}] \nl
SPS1312$-$0124\#007 & 13:11:46.32 & $-$02:35:33.8 & 23.58 & 0.792 & [O~{\sc iii}], H$\beta$ \nl
SPS1150+1242\#005 & 11:49:58.99 & 12:42:42.8 & 22.56 & 0.787 & [O~{\sc iii}], H$\beta$ \nl
SPS1514+3632\#008 & 15:14:22.68 & 36:31:31.1 & 22.84 & 0.801 & [O~{\sc iii}], H$\beta$ \nl
SPS1514+3632\#007 & 15:14:19.72 & 36:31:25.3 & 22.75 & 0.371 & [O~{\sc iii}], H$\beta$, [O~{\sc i}] \nl
SPS1550+2122b\#004 & 15:49:35.26 & 21:21:54.7 & 22.84 & 0.486 & [O~{\sc iii}], H$\beta$, [O~{\sc ii}] \nl
SPS1150+1242\#007 & 11:49:58.41 & 12:41:55.7 & 22.75 & 0.130 & H$\alpha$, [S~{\sc ii}], [O~{\sc i}] \nl
SPS1550+2122a\#008 & 15:49:36.15 & 21:21:45.4 & 23.61 & 0.573 & [O~{\sc iii}], H$\beta$, [O~{\sc ii}], H$-\delta$ \nl
\enddata
\end{deluxetable}
}

\clearpage

\begin{figure}

\plotone{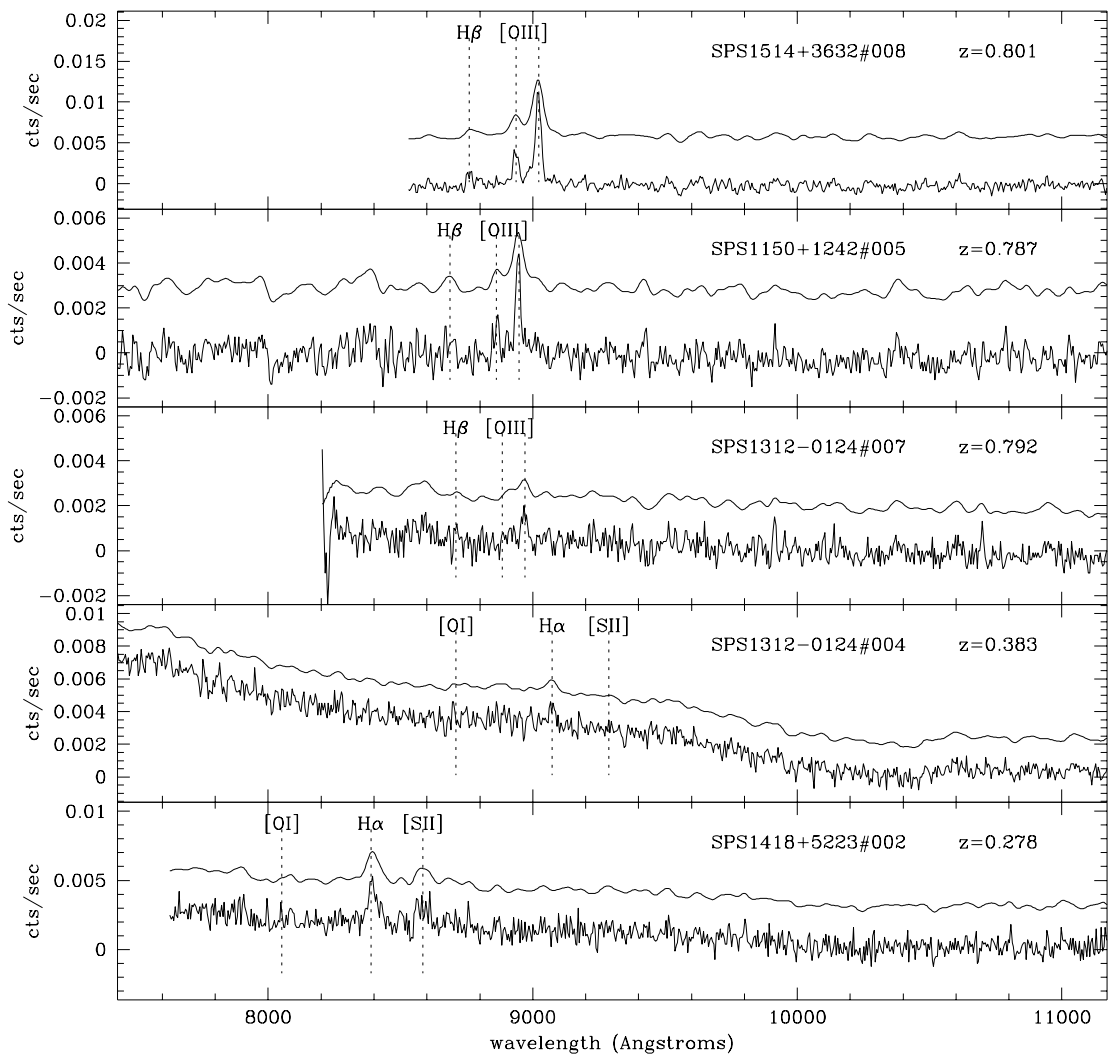}

\caption{Extracted spectra of emission line objects. In each panel,
the sky subtracted spectrum is shown, and a smoothed version of
the data is offset and plotted. Lines are marked at their laboratory 
wavelengths at the redshift of the object.}

\label{fig:spect.fig}

\end{figure}

\begin{figure}

\plotone{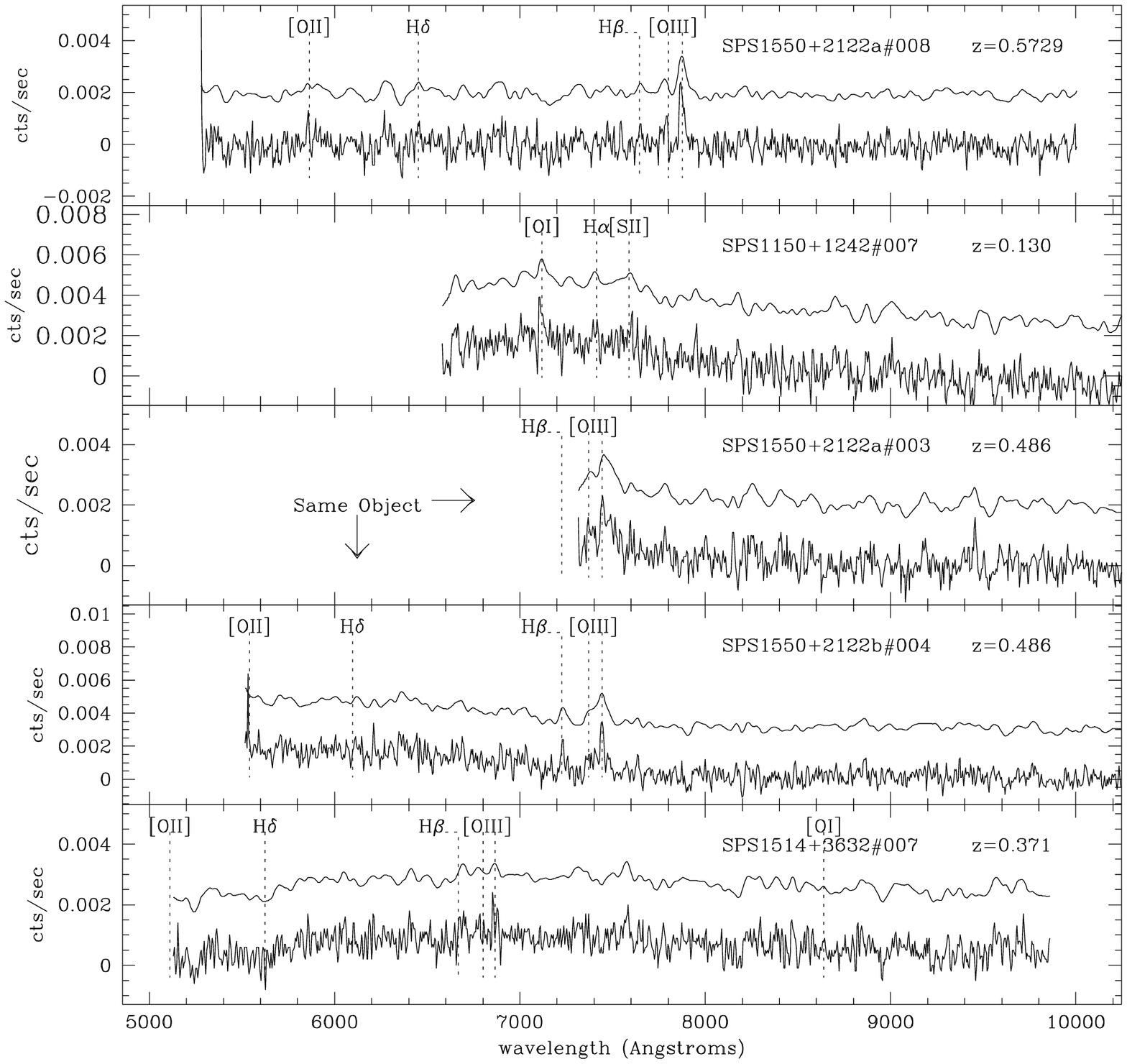}

\figurenum{1}

\caption{continued.}

\end{figure}

\begin{figure}

\vspace*{-1.0in}
\plotone{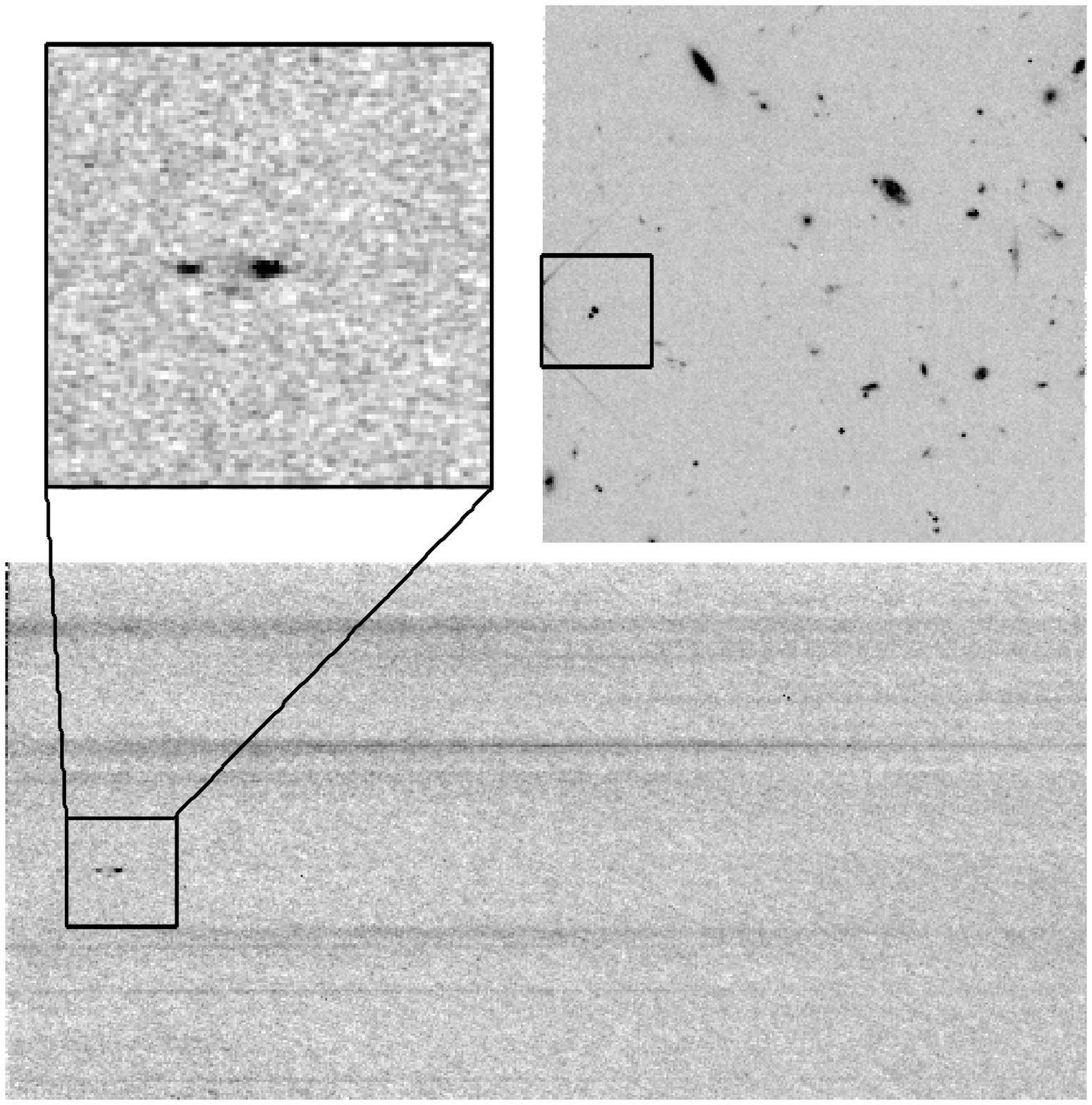}

\caption{The camera mode image, the spectral mode image and an
expanded region containing an emission line galaxy from SPS1514+3632.
Two galaxies, possibly an interacting pair, are seen at z=0.801,
and the [O~{\sc iii}] doublet is clearly seen for the brighter
object.}

\label{fig:image.fig}

\end{figure}

\end{document}